\documentclass[aps,prl,twocolumn,showpacs]{revtex4}
\input epsf
\begin{document}

\title{ Ab-initio simulation of photoinduced transformation of small rings in amorphous silica}
\author{Davide Donadio}
\affiliation{Computational Science, Department of Chemistry and Applied Biosciences, ETH Zurich
USI Campus, Via Giuseppe Buffi 13, CH-6900 Lugano, Switzerland}
\author{Marco Bernasconi}
\affiliation{Dipartimento di Scienza dei Materiali and INFM,
 Universit\`a di Milano-Bicocca, via Cozzi 53, I-20125
Milano, Italy}

\begin{abstract}
\noindent 
We have studied the photoinduced transformation of small rings (3-membered) in amorphous silica by
Car-Parrinello simulations. The process of ring opening leading to the formation of a couple of 
paramagnetic centers, namely an E' and a non-bridging-oxygen hole center (NBOHC), has been proposed 
experimentally to 
occur in silica exposed to  F$_2$  laser irradiation (at 7.9 eV). By  using a new scheme for the 
simulation of rare events in ab-initio molecular dynamics
(Iannuzzi, Laio and Parrinello, Phys. Rev. Lett. {\bf 90}, 238303 (2003)), we have identified the  transformation 
path for the opening of a 3-membered ring induced by a self-trapped triplet exciton, the 
migration of NBOHC and formation of a 
couple of stable E' and  NBOHC paramagnetic defects.

\end{abstract}
\maketitle 

\noindent

Extensive experimental and theoretical studies have been devoted to point defects in amorphous silica due to 
their relevance in the degradation of SiO$_2$-based electronic devices and in the photosensitivity of optical 
fibers \cite{pacchioni}. For instance, it is well established that the 
possibility to change the refractive index and write Bragg gratings in optical fibers by UV 
illumination  is connected to the photoinduced transformation of pre-existing defects or the formation of new 
ones.  The formation of defects in amorphous silica upon F$_2$ excimer laser irradiation is also held responsible 
for the degradation of lenses for optical lithography in semiconductor technology \cite{hosono}.

   In a recent experimental work Hosono {\sl et al.}
   \cite{hosono}  proposed that the main channel for color center 
   formation in silica exposed to radiation of an F$_2$ excimer laser (7.9 eV, well below the Tauc band gap of
   a-SiO$_2$)  is the generation of  an E' 
   center and  a non-bridging-oxygen hole center (NBOHC)  from one-photon excitation and breaking of strained bonds in
    small (3- or 4-membered) rings. The experimental evidence comes from combined 
   Electron Paramagnetic Resonance (EPR), Raman and optical spectroscopy measurements which show a 
   correlation between the intensities of the Raman lines associated to the breathing modes of 3- and 
   4-membered rings 
   ($D_1$ and $D_2$ lines)
   in samples with different fixation temperatures and the appearance of EPR and optical 
   absorption signals assigned to E' and NBOHC radicals.
   From these correlations Hosono {\sl et al.} inferred that one-photon absorption processes at $\sim$7.9 eV 
   generate excitons self-trapped on the small rings which would lead to ring opening and pair formation of
   E' and NBOHC defects as shown in Fig. 1.

\begin{figure}
\begin{center}
\epsfxsize= 8.5 truecm
\centerline{\epsffile{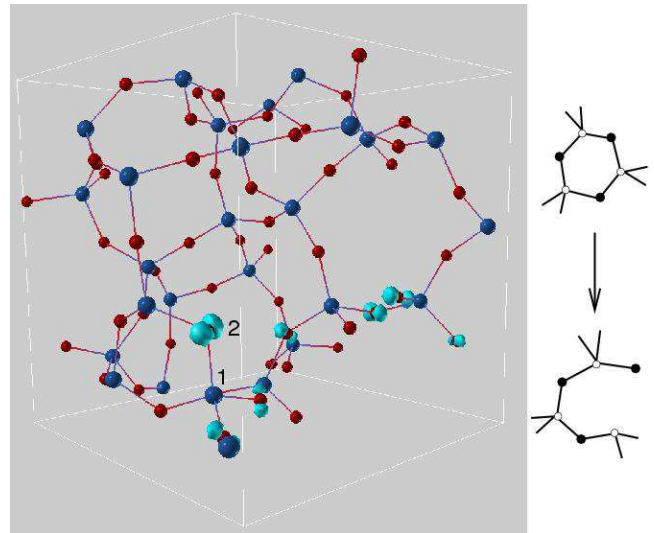}}
\caption{
Left panel: electron  density of the hole of the triplet exciton self-trapped on a 3-membered 
ring in a 81-atoms model of a-SiO$_2$. The contour value on the electron density plot is 0.125 a.u..
The Kohn-Sham orbital is mostly localized on the oxygen atom O(2).
Right panel: path for the pair generation  of the E' and NBOHC radicals upon photoexcitation of a 
3-membered ring as proposed in Ref. \protect\cite{hosono}.}
\end{center}
\end{figure}

   In this paper we report the results of ab-initio simulations of the photoinduced processes leading
   to the opening of small rings in a-SiO$_2$ aiming at providing theoretical support to the
   mechanism of E'-NBOHC pair generation proposed experimentally. The main outcome of the simulation
   is that the configuration of the two paramagnetic centers  proposed experimentally and shown in  
   Fig. 1  is  metastable on the excitate state, but the ring   spontaneously closes  once the
   system is brought back on the electronic ground state. However, a stable E'-NBOHC pair is found via
   a further migration of the NBOHC. 

   Ab-initio Car-Parrinello \cite{cpmd,CPMD,CPMD1} molecular 
   dynamics (MD) simulations have been performed within the framework of
   density functional theory in the local spin density approximation supplemented by generalized gradient corrections
   \cite{becke,lyp}. Norm conserving pseudopotentials \cite{troullier}, plane wave expansion of Kohn-Sham (KS) orbitals
   up to a kinetic energy cutoff of 70 Ry, a fictitious electronic mass of 800 a.u. and a time step of 0.15 fs have been used.
   The photoinduced transformations have been simulated   by adiabatic Born-Oppenheimer MD on the lowest 
   triplet (T$_1$) excited  state energy surface. 
 Although the photoinduced reaction would occur on the singlet excited state, MD on the T$_1$ excited  energy surface
 is more easily affordable within our framework and would provide 
  transformation paths that also  shed light on
  the processes induced by the singlet excitons as we have already demonstrated in a previous work on the
  photoinduced interconversion of oxygen deficient centers in a-SiO$_2$ \cite{odc}.

Periodic models of  a-SiO$_2$ containing 81 atoms at the experimental density of 2.2 g/cm$^3$
have been generated by quenching from the melt in classical MD, adopting the 
empirical pair potential by van Beest {\sl et al.} \cite{vanbeest}.
Quenching times as long as 5 ns have been used in
the classical simulations in order to generate models with a small number of 3- and 4-membered rings.
Three models have been generated, one with one 3- and four 4-membered rings, a second 
with one 3- and five 4-membered rings, and a third one with no 3- and one 4-membered ring.
The classical models have been then annealed  at 600 K for 1.5 ps 
by ab-initio MD. This procedure has been already demonstrated to provide models of a-SiO$_2$ with good
structural, elastic and dielectric properties \cite{pockel,benoit,pasquarello}.
By exciting the system on the lowest triplet state,  
we have found that for the first  among 
 the models  considered the hole of the exciton is partially localized on the 3-membered ring. 
 The self-trapped triplet exciton is shown in Fig. 1.
 Hereafter all the simulations  presented would refer to this latter model.
 The triplet exciton induces a slight lengthening of the Si(1)-O(2) bond in Fig. 1  by 0.03 \AA,  but the ring
 is still locally stable. The transformation leading to the opening of the ring turns out to be an activated
 process with an energy barrier much larger than the thermal energy. Thus it would not occur
 spontaneously on the time scale of Car-Parrinello simulations, few tens of ps long.
In order to overcome this limitation, we have exploited a new technique \cite{ilp} recently devised to 
simulate rare events  
within Car-Parrinello MD. The method is based on a coarse-grained dynamics in the 
space of few reaction coordinates, biased by a history-dependent potential which drives the system toward the lowest 
transition state \cite{ilp,laio,micheletti}. The total energy of  the 
locally stable configurations and  transition states found along the 
transformation path have been  then refined by constrained geometry optimization \cite{blumoon}.
The activation energies of the different processes have been then estimated,  although 
in principle also the entropic contributions to the activation free energies could be computed within the
method of Ref. \cite{ilp}. Due to the relatively large size of our simulation cell and the long simulation time
needed to get accurate free energy estimates, we have  restricted ourselves to the calculation of activation energies
and not free energies. Entropic effects are  small anyway  for the   
breaking of  siloxane bonds at room temperature as shown in Ref. \cite{boero} .

Following the scheme of Ref. \cite{ilp}, the collective  reaction coordinates
$S_{\alpha}(\{{\bf R}_I\})$, function of the ionic positions ${\bf R}_I$, define
a set of associated collective  variables $s_{\alpha}$
 which are treated as new dynamical variables. The extended system
is described by the Lagrangian

\begin{eqnarray}
L = &L_o& +
 \sum_{\alpha}\frac{1}{2}M_{\alpha}\dot{s}_{\alpha}^2
-\sum_{\alpha}\frac{1}{2}k_{\alpha}\bigl(S_{\alpha}(\{{\bf R_I}\})-
s_{\alpha}\bigr)^2 \\ \nonumber
&-&V(t,\{s_\alpha\}),
\label{lagran}
\end{eqnarray}

where $L_o$ is the Car-Parrinello Lagrangian, the second term is the fictitious kinetic energy of 
the $s_{\alpha}$'s, the third term is a harmonic potential that restrains the value of 
the collective coordinates  $S_{\alpha}(\{{\bf R_I}\})$ to the
corresponding dynamic collective variables $s_{\alpha}$. 
$V(t,\{s_\alpha\})$ is the Gaussian-like history dependent potential defined in ref. \cite{ilp}.

In a first simulation,
we have used  a single collective variable 
which provides a measure of the length of the Si(1)-O(1) bond (cfr. Fig. 1) which is lengthened upon excitation on
the triplet state. Namely, we have chosen the coordination number  $n_{Si(1)-O(2)}$ between atoms Si(1) and O(2).
Following Ref. \cite{ilp}, the coordination number between atom $a$ and atoms $b$ is defined as
$n_{a}=\sum_b \frac{ 1-(\frac{r_{ab}}{d})^6 }{ 1-(\frac{r_{ab}}{d})^{14} }$ where $r_{ab}$ is the
distance between the two atoms  and  scaling factor $d$ is chosen as $d$=2.2 \AA.
$n_a$ estimates the number of atoms $b$ within the bond cutoff distance to atom $a$
and decays smoothly for larger distances.
Recent investigations have shown that the best efficiency of the method would be achieved in our case 
when the collective variable has 
a characteristic frequency ($\sqrt{k/M}$)  
of the same order of magnitude as the Si-O stretching mode 
\cite{laiopriv}. 
Accordingly,  we have   chosen $M$= 1.82 10$^5$ a.u. and $k$= 1 a.u.   
in Eq. (1) ($\sqrt{k/M}$ $\sim$ 10$^{14}$ s$^{-1}$).
The parameters which  define the history-dependent
Gaussian potential in Eq. (3) of Ref. \cite{ilp} are $\delta s^{\perp}=0.08$ and $W=0.002$ Hartree.  
MD simulations have been performed at 300 K \cite{thermostat}.

\begin{figure}
\begin{center}
\epsfxsize= 8.5 truecm
\centerline{\epsffile{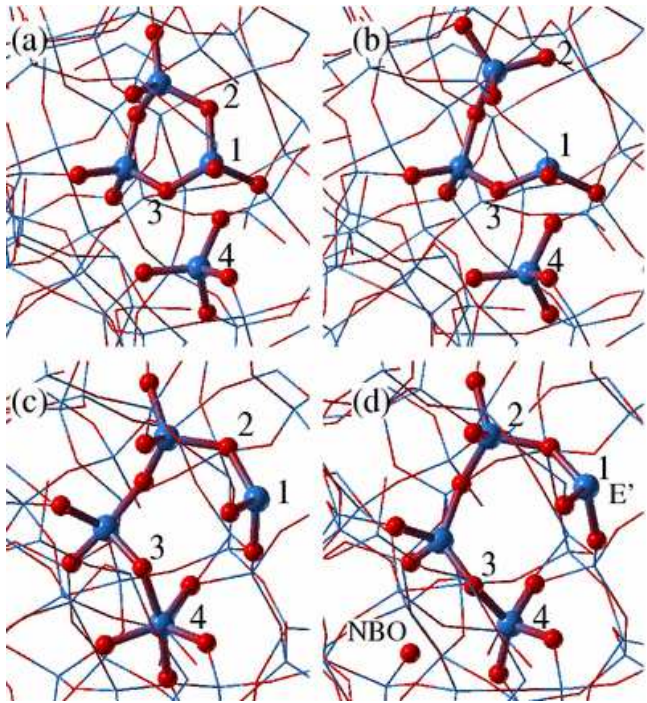}}
\caption{
Snapshots of the locally stable configurations found along the path that leads to the formation of 
an E'-NBOHC pair from the photoinduced opening of a 3-membered ring. The atoms involved in the transformation are 
represented by spheres and connected by thicker sticks. (a) The 3-membered ring excited on the T$_1$ state.
(b) The adjacent E'-NBOHC defect pair.  (c) The 
 E' center and the 5-coordinated silicon atom. (d) The stable E'-NBOHC pair.}
\end{center}
\end{figure}

\begin{figure}[t]
\begin{center}
\epsfxsize= 8.5truecm
\centerline{\epsffile{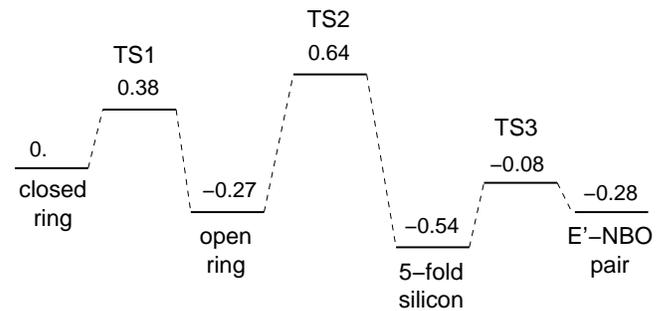}}
\caption{Energies of the optimized local minima corrsponding to the configurations of Fig. 2.
Closed ring, open ring, 5-fold silicon and E'-NBOHC pair refer to panels a, b, c and d of Fig. 2, respectively.
The energy and geometry of the transition states TS1, TS2 and TS3 are obtained from 
constrained geometry optimization \protect\cite{blumoon}
by using as reaction coordinate respectively the distance Si(1)-O(2), Si(1)-O(3) or Si(4)-NBO in Fig. 2.}
\end{center}
\end{figure}

Under the effect of the driving history-dependent potential the ring opens in 4.5 ps leading to the
configuration reported in Fig. 2b. 
Its energy, obtained by geometry optimization, is -0.27 eV with respect
to the energy of the closed ring on the T$_1$ excited state which is chosen hereafter as our zero of energy.
At this point the electron of the triplet exciton is self-trapped on Si(1).
In the  configuration of Fig. 2b  
two paramagnetic defects are formed (an E' center on Si(1) and a NBOHC on O(2), 3 \AA\ far apart) 
 by breaking the
Si(1)-O(2) siloxane bond and overcoming an activation barrier of 0.38 eV. The energy barrier  is obtained from constrained 
ab-initio MD simulations \cite{blumoon} at different values of a reaction coordinate chosen as the Si(1)-Si(2) distance.
The configuration in Fig. 2b is locally stable on the T$_1$ excited state, but 
by de-exciting the system  on the ground state (S$_o$), we have observed a charge transfer from the E'
center to the NBOHC which leads to a spontaneous closure of the ring \cite{notaB3lyp}. 
Therefore, the  two paramagnetic defects in the 
configuration of Fig. 2b are not stable in the ground state and cannot be assigned to the EPR 
signals observed experimentally. 
We may envisage to stabilize  the E'-NBOHC pair by reducing the interaction between the two
radicals to be realized by allowing a 
migration of one of two paramagnetic centers.  In order to identify this process we have added  a second collective variable
in Eq. (1),  defined as the coordination number $n_{Si(1)-O(3)}$ between atoms Si(1) and O(3) in Fig. 2a.
The presence of both the $n_{Si(1)-O(2)}$ and $n_{Si(1)-O(3)}$ collective variables  allows to explore
different  paths for the ring opening. 
The parameters $M$,  $k$, $\delta s^{\perp}$ and $W$ are the same as before.
In this second simulation the system initially follows the same path 
seen in  the previous one  which means that the
Si(1)-O(2) bond is weaker than the Si(1)-O(3) one.
However, as the simulation proceeds,  the history-dependent potential drives the system away 
from the local miminum of Fig. 2b.  
First,  the Si(1)-O(2) bond is  recovered and then  a new conformation of the 
open ring is formed by  breaking  the Si(1)-O(3) bond. 
The NBOHC generated in this way (on O(3)) 
floats around and finally forms a new bond with Si(4) which becomes 5-fold coordinated. 
This process has a higher activation barrier (0.64 eV) but 
leads to a configuration (Fig. 2c) which is locally stable  with a  total energy of -0.54 eV.  
From this configuration it is now envisageable to generate a new NBOHC 
by breaking one of the five Si(4)-O bonds. To identify which Si(4)-O bond is most prone to break, we have started 
a third simulation from the configuration of Fig. 2c  with the total coordination number $n_{Si(4)-O}$ 
as collective variable.
The final structure  of this latter process  is shown in Fig. 2d. Its total energy is -0.28 eV and the
activation barrier for the formation of the NBOHC is 0.46 eV (Fig. 3). 
The E' and the NBOHC centers in this final configuration are distant enough (6.3 \AA)
to prevent the  charge transfer 
upon de-excitation on the S$_o$ state. As the two radicals do not interact  each other, there is no singlet-triplet 
splitting and the EPR signals are supposed to be equal to those of the isolated paramagnetic centers.
The path identified by the local minima of Figs. 2a-c  represents thus a viable mechanism for the opening of the ring
and migration of the NBOHC suitable to provide a pair of stable paramagnetic defects. 
The total trajectory starting from the initial configuration of Fig. 2a and reaching the final configuration
of the stable E'-NBOHC pair in Fig. 2d is 2.7 ps long .
The energies of the configurations
in Fig. 1 and of the transition states separating the different local minima are collected in the scheme of Fig. 3.
The geometry of  the transition state leading to the 5-fold coordinated Si is reported in Fig. 4.
We remark that the recovery of the Si(1)-O(2) bond moving from configuration in Fig. 2b to the 
transition state in Fig. 4 does not restore the initial configuration of the closed ring (Fig. 2a). No
intermediate local minima from Fig. 2b to Fig. 2c is found. 
The activation barrier for the whole process (the energy of the highest transition state) is 0.64 eV, sufficiently low
to make this transformation channel viable at room temperature. 
These values for the activation energies are supposed to depend on the local environment and strain of the
small ring. As a consequence only a fraction of the small rings are expected to undergo this trasformation
upon photoexcitation.

\begin{figure}
\begin{center}
\epsfxsize= 5.truecm
\centerline{\epsffile{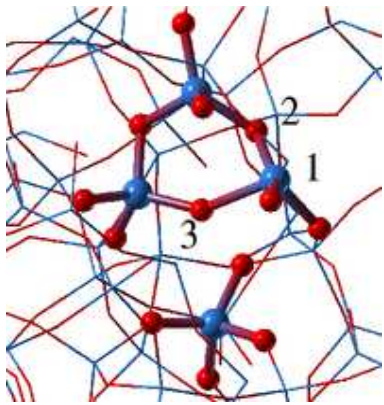}}
\caption{Geometry of the transition state TS2 of Fig. 3. The distances between the silicon
atom Si(1)  and the oxygen atoms O(2) and O(3) are 1.80 \AA\ and 2.05 \AA, respectively. The tetrahedron
formed by atom Si(1) is highly distorted and undergoes a puckering 
 once a NBOHC is formed on O(3) (cfr. Fig. 2c). As a consequence the
resulting E' center that does not point towards the NBOHC.}
\end{center}
\end{figure}

   In summary, we have identified the transformation path for the opening of a 3-membered ring induced by a 
   self-trapped triplet exciton. Once open in the excited electronic state, the 3-membered ring gives rise to a 
   couple of strongly interacting E' and NBHOC defects, which recombine quickly after electronic de-excitation. 
   However, a simple path with a low energy barrier (0.64 eV) has been identified for the migration of the NBOHC which 
   leads to the formation of a couple of stable E' and NBOHC paramagnetic centers. The simulations thus provide a 
   theoretical support to the interpretation of the experimental data 
   in Ref. \cite{hosono} on the formation of color centers in a-SiO$_2$ by F$_2$ laser irradiation.

We gratefully thank M. Iannuzzi, A. Laio and M. Parrinello for
discussion and for sharing with us their insight on the new methods they have developed.
D.D. aknowledges financial support from Pirelli Cavi e Sistemi S.p.a.
This work is partially supported by the INFM Parallel Computing Initiative. 

$^*$corresponding author: ddonadio@phys.chem.ethz.ch


\begin{thebibliography}{54}
\bibitem{thermostat} Constant temperature is enforced by velocity rescaling at the target
temperature with a tolerance of 10 $\%$.
\bibitem{notaB3lyp} We have performed additional calculations on cluster model by using
the B3LYP \protect\cite{lyp,B3LYP} functional in order to check the effect of the functional on the
E' $\rightarrow$ NBOHC  charge transfer mechanism identified in the simulation.
In fact, the BLYP functional is expected to underestimate
the Hubbard-U parameter on the oxygen radical due to incomplete correction of the self-interaction.
On the contrary the self-interaction is exactly cancelled within the B3LYP functional. We have considered two clusters
$\cdot$Si(OSiH$_3$)$_3$ and $\cdot$O(SiH$_3$) at different Si-Si distances. We have found that  the
charge transfer occurs for E'-NBOHC distances below 5 \AA\ for the
B3LYP functional and below 5.4 \AA\ for the BLYP functional.
The calculations have been performed with the code Gaussian98 \protect\cite{Gaussian}
and  the 6-311+G$^*$ basis set which  provides a good description of
point defects in silica \cite{pacch99}.
\end{thebibliography}
\end{document}